\documentclass{article}
\usepackage{graphicx}  
\usepackage{amsmath}   
\usepackage[compress]{cite}
\usepackage{amssymb}   
\usepackage{bm} 
\usepackage{dcolumn}
\usepackage{color}
\usepackage{mathrsfs}
\usepackage{amsfonts}
\usepackage{varioref}
\usepackage{textcomp}
\RequirePackage[colorlinks,citecolor=blue,urlcolor=magenta,linkcolor=blue]{hyperref}
\allowdisplaybreaks
\labelformat{section}{Section #1} 
\labelformat{subsection}{Section #1} 
\labelformat{subsubsection}{Section #1}
\labelformat{subsubsubsection}{Section #1}
\labelformat{equation}{Eq.~(#1)} 
\labelformat{figure}{Fig.~#1} 
\labelformat{subfigure}{Fig.~\thefigure#1} 
\labelformat{table}{Tab.~#1} 
\labelformat{appendix}{Appendix #1}
\title{Boundary Term in the Gravitational Action is the Heat Content of the Null surfaces}
\author{Sumanta Chakraborty\footnote{sumantac.physics@gmail.com}~
$^{1}$ and T. Padmanabhan\footnote{paddy@iucaa.in}~$^{2}$\\
{$^{1}$\small{School of Mathematical and Computational Sciences and School of Physical Sciences}}\\
{\small{Indian Association for the Cultivation of Science, Kolkata-700032, India}}\\
{$^{2}$\small{IUCAA, Post Bag 4, Ganeshkhind, Pune University Campus, Pune 411007, India}}}
\begin{document}
  
\maketitle
\begin{abstract}

The Einstein-Hilbert Lagrangian has no well-defined variational derivative with respect to the metric. This issue has to be tackled by adding a suitable surface term to the action, which is a peculiar feature of gravity. We also know that null surfaces in spacetime exhibit (observer-dependent) thermodynamic features. This suggests a possible thermodynamic interpretation of the boundary term when the boundary is a null surface. For timelike/spacelike surfaces it is easy to construct the boundary term but there are some subtleties in the case of the null surface. The correct form of boundary term for null surfaces was obtained recently from first principles. We show that this surface term, as well as its variation, have direct thermodynamic interpretation in terms of a heat density of null surfaces. The implications of the result are discussed.

\end{abstract}
\section{Introduction}

The standard action principle in general relativity, based on the Lagrangian density $R\sqrt{-g}$ does not lead to a well defined variational principle unless we take care of the boundary contributions in some suitable manner \cite{Padmanabhan:2010zzb,York:1972sj,Gibbons:1976ue,Charap:1982kn,Dyer:2008hb,Padmanabhan:2014lwa}. There are two possible ways of handling this situation. 

The first is to separate the action into a bulk term --- which is quadratic in the first derivatives of the metric --- and a boundary term arising from the second derivatives of the metric. Once this is done, one can simply discard the boundary term and work with the quadratic action. While the separation is foliation dependent the resulting field equations are not and everything works out satisfactorily. 

The second procedure is to add a suitable boundary term to the Einstein-Hilbert action and arrange matters such that the variation of the boundary term cancels the unwanted surface variation terms coming from the Einstein-Hilbert action. The resulting action principle will lead to the standard field equations if the 3-metric is fixed at the boundary. 

In either approach the nature of the boundary term --- arising from the Einstein-Hilbert action or due to the addition of the extra term --- depends on the nature of the boundary. For example, in the case of a space-like or time-like boundaries, the additional term can be built from the trace of the extrinsic curvature \cite{Padmanabhan:2010zzb,York:1972sj,Gibbons:1976ue}. In the case of a null boundary, or in those parts of the boundary which are null, one cannot define the extrinsic curvature in a natural manner.  It can, however, be shown that in the case of null boundaries the surface term is essentially the sum $\Theta + \kappa$ where $\Theta$ is the expansion and $\kappa$ is the surface gravity of the null congruence defining the boundary surface \cite{Parattu:2015gga,Parattu:2016trq,Chakraborty:2016yna,Lehner:2016vdi,Jubb:2016qzt,Feng:2017ygy,Wieland:2017zkf}. 

The existence of the null surfaces is a unique feature of gravitational theory. Acting as one way membranes, the null surfaces can restrict access to spacetime events for  a class of observers confined to one side of the surface. Familiar examples include event horizons in black hole spacetimes, de Sitter spacetime and the case of uniformly accelerated observers in flat spacetime. In all these cases, observers who perceive a horizon attribute to it a temperature $T$. Further, the lack of accessibility of spacetime events beyond the horizon suggests attributing an entropy density $s$ to the null surface. Together, this allows us to associate a heat density $Ts$ to any patch of null surface \cite{Gibbons:1977mu,Iyer:1994ys,Padmanabhan:2007en,Padmanabhan:2009vy,Padmanabhan:2014jta,Chakraborty:2015hna,Majhi:2013jpk}. 

Such an association of thermodynamic variables with spacetime finds a natural backdrop in the emergent gravity paradigm in which one attempts to describe gravity in a manner analogous to, say, fluid dynamics or elasticity. In this approach gravitational dynamics is not described in terms of geometrical concepts but instead is phrased completely in thermodynamic language.\footnote{Note that this is conceptually very different from attempts to derive gravitational field equations from thermodynamic considerations, finally obtaining $G_{ab} = 8\pi T_{ab}$  with the left hand side still being treated as having a geometric interpretation. In contrast, the emergent gravity paradigm interprets the field equation entirely in thermodynamic language.} The evolution of geometry in a bulk region of space, for e.g., is described in terms of heating/cooling of the thermodynamic degrees of freedom driven by the disparity between the degrees of freedom on the surface and bulk regions of spacetime \cite{Padmanabhan:2013nxa,Padmanabhan:2010xh,Padmanabhan:2010rp,Parattu:2013gwa,Chakraborty:2014rga,Chakraborty:2015aja}. 

The existence of such an interpretation suggests that even the conventional approach to gravity should exhibit traces of its underlying thermodynamic character. In particular, we would expect any peculiar feature of gravitational action principle --- not usually found in other field theories --- to be connected with the fact that gravitational dynamics has an underlying thermodynamic interpretation. As we noted before, a striking feature of gravitational dynamics is the existence of a surface term in the action principle. This fact, coupled with the second peculiar feature of gravity viz. the existence of null surfaces which can act as one way membrane, suggests that there must exist a direct connection between the surface term in the action principle and the thermodynamics of null surfaces \cite{Mukhopadhyay:2006vu,Padmanabhan:2010xe}. The purpose of this paper is to establish this result. (Such a connection was established in some of the previous works (see e.g. \cite{Majhi:2013jpk}), treating the null surface as a limiting case of timelike surfaces. Recently, we provided a first principle derivation of the boundary term for the null surface, without using any limiting procedure \cite{Parattu:2015gga,Parattu:2016trq}. Here we will not use any limiting procedures but instead will use this more rigorous approach. Further, unlike previous works, our result will be \textit{completely general} and we will not require any special assumptions regarding the nature of the spacetime or other approximations.)

We will show that there exists a simple but at the same time very general interpretation of the surface term in the Einstein-Hilbert action evaluated on a null surface and the heat density of the null surface. When the null surface is described using Gaussian null coordinates, the boundary term in the action reduces to the heat density $Ts$ of the null surface. Moreover, the variation of the surface term arising from a ``flow'' along the null congruence induced by the transformation $x^a \to x^a + \ell^a$ can be expressed in the form of the variation $T\, \delta s$, once again providing a thermodynamic condition. These results provide yet another link between the gravitational dynamics and spacetime thermodynamics in the context of null surfaces.

In what follows we will use the following mostly positive signature convention for the metric and shall set the fundamental constants $c$ and $G$ to unity. The Roman indices $a,b,\ldots$ will stand for four dimensional spacetime coordinates and Greek indices $\mu,\nu,\ldots$ will stand for spatial coordinates.

The rest of the paper is organized as follows: We start in \ref{Sec_02} by computing the boundary contribution of the total gravitational action associated with null surfaces and exploring its thermodynamic properties. Subsequently in \ref{Sec_03} we have discussed a certain variation of the null boundary term having a nice thermodynamic interpretation. Finally we conclude with a discussion on our results. Relevant calculations are presented in the four appendices, \ref{AppA}, \ref{AppGNC}, \ref{AppB} and \ref{AppC}, respectively.

\section{Boundary term in the action is the Heat content}\label{Sec_02}

Our first task will be to show that the boundary term which one must add to the Einstein-Hilbert action, when evaluated within a volumed bounded by null surfaces, in order to get rid of the second derivatives of the metric, has a natural thermodynamic interpretation as the heat density of the null boundaries. For this purpose, we express the Einstein-Hilbert action separated into a bulk and a boundary term. Further, in order to uncover the thermodynamic nature of the underlying gravitational dynamics using the boundary term, it is convenient to work with the following variables \cite{Parattu:2013gwa}:
\begin{equation}
f^{ab}=\sqrt{-g}g^{ab};
\qquad 
N^{c}_{ab}=-\Gamma^{c}_{ab}+(1/2)(\delta ^{c}_{a}\Gamma ^{m}_{mb}+\delta ^{c}_{b}\Gamma ^{m}_{ma})
\end{equation} 
In terms of these variables, the Einstein-Hilbert action separates neatly into a bulk and a boundary term:
\begin{align}\label{Therm_01}
\sqrt{-g}R=\frac{1}{2}N^{c}_{ab}\partial _{c}f^{ab}+\partial _{c}\left(-f^{ab}N^{c}_{ab}\right)
\equiv \sqrt{-g}L_{\rm bulk} +\sqrt{-g}L_{\rm sur}
\end{align}
Thus one can immediately identify the term $\partial_{c}(f^{ab}N^{c}_{ab})$ as the boundary term one must add to the Einstein-Hilbert action, which when integrated over a boundary surface $\phi=\textrm{constant}$, yields $s_{c}f^{ab}N^{c}_{ab}$ with normal $s_{c}=\nabla_{c}\phi$. In what follows we will concentrate on the structure of this boundary term, in the context of null surfaces (For a brief discussion on the corresponding situation for spacelike/timelike surfaces we refer the reader to \ref{AppA}).  

Thus our job is to start from the boundary term $s_{c}f^{ab}N^{c}_{ab}$, one must add to the Einstein-Hilbert action, integrated over the boundary surface and then considering the thermodynamic interpretation as the boundary surface becomes null. In this spirit, we shall first rewrite the boundary term in a different manner, valid for an arbitrary surface and then we will specialize to the null surface. For clarity, we present below the full structure of the boundary term,
\begin{align}
\mathcal{S}=\frac{1}{16\pi G}\int_{\partial \mathcal{V}} d^{3}x~s_{c}f^{ab}N^{c}_{ab}~,
\end{align}
where, as mentioned earlier $\partial \mathcal{V}$ is the $\phi=\textrm{constant}$ surface and $s_{c}=\nabla_{c}\phi$ is the normal to the desired boundary surface, which we keep arbitrary for the moment. By expanding out the Christoffel symbol, the integrand can be rewritten as:
\begin{align}
s_{c}f^{ab}N^{c}_{ab}
=f^{ab}\left[-s_{c}\Gamma^{c}_{ab}+\frac{1}{2}s_{a}\Gamma^{d}_{bd}+\frac{1}{2}s_{b}\Gamma^{d}_{ad}\right]~.
\end{align}
The product of the normal vector and the Christoffel symbols can be expressed in terms of covariant derivatives of the normal vector, in particular, we can use the following results, 
\begin{align}
\nabla_{a}s_{b}=\partial_{a}s_{b}-\Gamma^{c}_{ab}s_{c};\qquad \nabla_{a}s^{a}=\partial _{a}s^{a}+\Gamma^{a}_{ac}s^{c}~,
\end{align}
which along with the definition $f^{ab}=\sqrt{-g}g^{ab}$ can be used to express the boundary term as, 
\begin{align}
s_{c}f^{ab}N^{c}_{ab}&=\sqrt{-g}g^{ab}\left(\nabla_{a}s_{b}-\partial_{a}s_{b}\right)+\sqrt{-g}\left(\nabla_{a}s^{a}-\partial _{a}s^{a}\right)
\nonumber
\\
&=\sqrt{-g}\left(2\nabla_{a}s^{a}-g^{ab}\partial_{a}s_{b}-\partial _{a}s^{a}\right)~.
\end{align}
Note that the boundary term is not covariant, since it explicitly depends on the Christoffel symbol.\footnote{This is a feature and not a bug. We expect null surfaces in flat spacetime to exhibit thermal properties; so the action cannot vanish in flat spacetime if we use non-inertial coordinates. The correct action for gravity --- unlike the Einstein-Hilbert action --- satisfies this criterion.} The above expression is another manifestation of the same, since it depends on the partial derivatives of the normal vector. To cast it into the desired form, it is advantageous to introduce the following projector $\Pi^{a}_{~b}=\delta^{a}_{b}+t^{a}s_{b}$ \cite{Parattu:2016trq}, where $t_{b}$ is an auxiliary vector satisfying $s_{a}t^{a}=-1$, such that $s_{a}\Pi^{a}_{~b}=0$. Use of this projector helps us to write down the boundary term in the following manner, 
\begin{align}
s_{c}f^{ab}N^{c}_{ab}&=2\sqrt{-g}~\left\{\Pi^{a}_{~b}\nabla_{a}s^{b}\right\}+\sqrt{-g}\left(-g^{ab}\partial_{a}s_{b}-\partial _{a}s^{a}\right)-2\sqrt{-g}t^{a}s_{b}\nabla_{a}s^{b}~.
\end{align}
The discussion so far is applicable to any boundary, irrespective of whether it is null or spacelike/timelike. Subsequently, we will specialize to the case of a null boundary for which we can take $s_{a}=\ell_{a}$ and $t_{a}=k_{a}$, with the following properties: $\ell_{a}\ell^{a}=0=k_{a}k^{a}$ and $\ell_{a}k^{a}=-1$. Thus for null boundaries, the surface term of the Einstein-Hilbert action becomes,
\begin{align}
s_{c}f^{ab}N^{c}_{ab}\big|_{\rm null}&=2\sqrt{q}~\left\{\Pi^{a}_{~b}\nabla_{a}\ell^{b}\right\}+\sqrt{q}\left(-g^{ab}\partial_{a}\ell_{b}-\partial _{a}\ell^{a}\right)-2\sqrt{q}k^{a}\ell_{b}\nabla_{a}\ell^{b}~.
\end{align}
In order to simplify the above expression and to attribute a thermodynamic meaning to this, it is convenient to go to a coordinate system where $\phi$ itself is a coordinate. This is very much similar to the Gaussian Null Coordinate system, which can be constructed for any arbitrary null surface, see \cite{Parattu:2015gga}. In this system of coordinates with $\ell_{a}=\nabla_{a}\phi$ we have, $k^{a}\ell_{b}\nabla_{a}\ell^{b}=-\kappa$, as well as $\Pi^{a}_{~b}\nabla_{a}\ell^{b}=\Theta+\kappa$. Further since $\phi$ itself is a coordinate, we immediately have $\partial_{a}\ell_{b}=0$. Thus the surface term becomes,
\begin{align}
s_{c}f^{ab}N^{c}_{ab}\big|_{\rm null}&=2\sqrt{q}\left(\Theta+\kappa\right)+2\kappa \sqrt{q}-\sqrt{q}\partial _{a}\ell^{a}~.
\end{align}
In order to evaluate the last term in this particular coordinate system, adapted to the null surface, we start with an alternative definition for $\kappa$, which is 
\begin{align}
\kappa=-\frac{1}{2}k^{a}\nabla_{a}\ell^{2}=\frac{1}{2}\partial_{\phi}g^{\phi \phi},
\label{nattemp}
\end{align}
where we have chosen $k^{\phi}=-1$, in order to satisfy the relation $\ell_{a}k^{a}=-1$. Further, $\partial_{a}\ell^{a}=\partial_{a}g^{a \phi}$ whose only non-zero component, on the null surface becomes, $\partial_{\phi}g^{\phi \phi}$. This is because derivative of all the other metric components vanish in the null limit (as is evident from the GNC parametrization, see \ref{AppGNC}). Thus we obtain the surface term on a null surface to be:
\begin{align}
\mathcal{S}=\frac{1}{16\pi G}\int _{\partial \mathcal{V}}d^{3}x~s_{c}f^{ab}N^{c}_{ab}\big|_{\rm null}
=\frac{1}{16\pi G}\int _{\partial \mathcal{V}}d^{3}x~2\sqrt{q}~\left(\Theta+\kappa\right)~.
\end{align}
We will now rewrite the first term in the integrand, namely $\sqrt{q}\Theta$ in the form $(d\sqrt{q}/d\lambda)$, since $\Theta=(d\ln \sqrt{q}/d\lambda)$. Thus integrating this expression we will obtain a  contribution at the end points $\lambda= \lambda_1$ and $\lambda= \lambda_2$ proportional to the areas of the 2-surfaces. Further we can  associate the temperature $T=(\kappa/2\pi)$ and  the entropy \textit{density} $s =(\sqrt{q}/ 4G)$, with the null surface. We then find that: 
\begin{equation}\label{Therm02}
\mathcal{S}=\int _{\partial \mathcal{V}}d^{3}x~Ts + \frac{1}{2\pi}(S_2-S_1)
\end{equation} 
where we have written the end point contributions in terms of the entropy $S=A/4G$ of the 2-surfaces at $\lambda= \lambda_1$ and $\lambda= \lambda_2$. We thus find that the boundary contribution to  the Einstein-Hilbert action from $\partial \mathcal{V}$ is indeed the integral of the heat density, $\mathcal{H}\equiv Ts$.

The end point contribution in \ref{Therm02}  vanishes in stationary situations with $A_2=A_1$ or if such bounary conditions are imposed. Incidentally, even when it is nonzero, it can be given a physical interpretation along the following lines: It is usual to take the dimensions of temperature to be $1/L$ and that of entropy \textit{density} as $(1/L^2)$ so that the heat density has the dimensions of $(1/L)(1/L^2)$; when integrated over  $\partial \mathcal{V}$, which provides a dimension $L^3$, we get the dimensionless action. The end point contribution, on the other hand, is obtained by an integral over $\partial \partial\mathcal{V}$ which provides the dimensions $L^2$ that is cancelled by the dimension $1/L^2$ of the overall constant $1/8\pi G$ in front of the integral. This means that, purely from dimensional perspective,  the end point contributions will be just the entropy while the term involving integral over $\partial \mathcal{V}$ will contain heat density in the integrand. To provide the same interpretation for both terms, we need to introduce a \textit{dimensionless} temperature. This will allow us  to interpret the end-point contribution as the \textit{heat} content (rather than the \textit{entropy} content) of the  2-surface $\partial \partial\mathcal{V}$. It is possible to do this because the numerical value, as well as the dimension of $\kappa$ can be changed by rescaling $\ell^a$. We have defined  $\kappa$  through the relation, $\ell^{a}\nabla_{a}\ell^{b}=\kappa \ell^{b}$, which depends on the parametrization of the null vector $\ell^{a}$. If we rescale $\ell^a\to f(x)\ell^{a}$, the surface gravity changes by $\kappa\to f(x)\{\kappa+\ell^{a}\nabla_{a}\ln f(x)\}$. Using this freedom one can always ensure the normalisation $\bar{\kappa}=1$, at the two end points $\lambda= \lambda_1$ and $\lambda= \lambda_2$. Then the temperature of the 2-surfaces $\partial \partial\mathcal{V}$ can be interpreted as being equal to $T=\kappa/2\pi =1/2\pi$, so that the end point contribution becomes $\Delta (TS)$ and we can write our result as:
\begin{equation}
\mathcal{S}=\int _{\partial \mathcal{V}}d^{3}x~Ts+\Delta (TS)
\end{equation} 
This interpretation is not essential to our result since we are only concerned with proving that contribution from $\partial \mathcal{V}$ is the integral of the heat density in the most natural context. However, it is interesting to note that the numerical factors work out correctly for us to identify $T=1/2\pi$ at the end points.

\section{Variation of the null boundary term: Thermodynamic interpretation}\label{Sec_03}

It turns out that not only the boundary term, but certain variation of the same has thermodynamic interpretation. In this section we will discuss such a variation and the associated interesting thermodynamic interpretation it presents. We would like to start by briefly mentioning the variation of the integrand of the Einstein-Hilbert action which yields, for arbitrary variations of the metric \cite{Parattu:2013gwa}:
\begin{align}\label{Therm_04}
\delta\left(\sqrt{-g}R\right)=-\partial _{c}\left(f^{ab}\delta N^{c}_{ab}\right)+R_{ab}\delta f^{ab}~.
\end{align}
Integrating over a spacetime volume the above variation of the action can be rewritten as:
\begin{align}\label{Therm_05}
\int_{\mathcal{V}} d^{4}x \delta \left(\sqrt{-g}R\right)=\int _{\mathcal{V}}d^{4}x R_{ab}\delta f^{ab}
-\int _{\partial \mathcal{V}}d^{3}x~s_{c}f^{ab}\delta N^{c}_{ab}~,
\end{align}
where the boundary surface $\partial \mathcal{V}$ is taken to be $\phi=\textrm{constant}$ with an unnormalised normal $s_{c}=\nabla _{c}\phi$. On the other hand, we can rewrite the Einstein-Hilbert Lagrangian density as, $\sqrt{-g}R=R_{ab}f^{ab}$, such that its variation becomes, $\delta (\sqrt{g}R)=R_{ab}\delta f^{ab}+f^{ab}\delta R_{ab}$. Hence \ref{Therm_05} suggests the following identity, $f^{ab}\delta R_{ab}=-\partial _{c}\left(f^{ab}\delta N^{c}_{ab}\right)$, which can also be derived starting from first principle computation \cite{Parattu:2013gwa}. 

As emphasized earlier, the boundary term in the variation of the gravitational action has the structure $s_{c}f^{ab}\delta N^{c}_{ab}$, where $s_{c}=\nabla_{c}\phi$ is the normal to the boundary hypersurface $\phi=\textrm{constant}$. We will now demonstrate that this variation also has a simple thermodynamic interpretation for variations of the metric arising out of the `flow' along the null congruence, such that $g^{ab}\to g^{ab}+(1/2)(\nabla^{a}\ell^{b}+\nabla^{b}\ell^{a})$. Arbitrary variations of the boundary term on a null surface can be computed in a straightforward manner, whose technical details have been delegated to \ref{AppB} and quote here the final result:
\begin{align}\label{Basic_Equation}
\delta \left\{\frac{1}{8\pi G}\int _{\partial \mathcal{V}}d^{2}xd\lambda\sqrt{-g}\left(\Theta+\kappa\right)\right\}
&=-\frac{1}{16\pi G}\int _{\mathcal{V}}d^{4}x~f^{ab} \delta R_{ab}
+\frac{1}{16\pi G}\int _{\partial \mathcal{V}}d^{3}x~\partial _{a}\left(\sqrt{q}\Pi^{a}_{b}\delta \ell^{b}\right)
\nonumber
\\
&+\frac{1}{16\pi G}\int _{\partial \mathcal{V}}d^{2}xd\lambda \sqrt{q} 
\Bigg[\left\{\Theta_{ab}-(\Theta+\kappa)q_{ab}\right\}\delta q^{ab}
+P_{a}\delta \ell^{a}\Big]~.
\end{align}
In the above expression, $\Theta_{ab}$ is the extrinsic curvature associated with the null generator $\ell_{a}$ and $\Theta$ is the trace of the extrinsic curvature. Further, the quantity $P_{a}$ has the following expression, $P_{a}=\big\{2k_{a}\left(\Theta+\kappa\right)-k^{b}\left(\nabla_{a}\ell_{b}+\nabla_{b}\ell_{a}\right) \big\}$, where $\kappa$ is the non-affinity parameter associated with the null generators and $k^{a}$ is the auxiliary null vector with $\ell_{a}k^{a}=-1$.

The above expression provided a general variation of the null boundary term for arbitrary variations. Our aim is to provide a thermodynamic interpretation for this variation and for this purpose it will be advantageous to consider variations associated with displacements along the null surface, leading to $\delta g^{ab}=(1/2)(\nabla^{a}\ell^{b}+\nabla^{b}\ell^{a})$, such that $\delta g_{ab}=-\nabla_{a}\ell_{b}$ (where we have used the result, $\ell_{a}=\nabla_{a}\phi$). For this variation induced by the flow along the null congruence we have: $f^{ab}\delta R_{ab}=-\sqrt{-g}\nabla _{a}(R^{a}_{b}\ell^{b})$, $\delta q^{ab}=\Theta^{ab}+\textrm{terms~proportional~to}(\ell^{a},k^{a})$ and $\delta \ell^{a}=\kappa \ell ^{a}$. Thus from \ref{Basic_Equation} we obtain,
\begin{align}
\delta \left\{\frac{1}{8\pi G}\int _{\partial \mathcal{V}}d^{2}xd\lambda\sqrt{-g}\left(\Theta+\kappa\right)\right\}
&=\frac{1}{16\pi G}\int _{\partial \mathcal{V}}d^{2}xd\lambda \sqrt{q} R_{ab}\ell^{a}\ell^{b}
+\frac{1}{16\pi G}\int _{\partial \mathcal{V}}d^{3}x~\partial _{a}\left(\sqrt{q}\Pi^{a}_{~b}\delta \ell^{b}\right)
\nonumber
\\
&+\frac{1}{16\pi G}\int _{\partial \mathcal{V}}d^{2}xd\lambda \sqrt{q} 
\Bigg[\left\{\Theta_{ab}\Theta ^{ab}-(\Theta+\kappa)\Theta\right\}
+\kappa P_{a} \ell^{a}\Big]~.
\end{align}
In the above expression, for the variation considered earlier, we have $\Pi^{a}_{~b}\delta \ell^{b}=\kappa \ell^{a}$ and hence we obtain, $\partial _{a}(\sqrt{q}\Pi^{a}_{~b}\delta \ell^{b})=(1/\sqrt{q})(d/d\lambda)(\kappa \sqrt{q})=(d\kappa/d\lambda)+\kappa \Theta$. The $R_{ab}\ell^{a}\ell^{b}$ term appearing in the above variation can be transformed to various geometric quantities associated with the null surface by using the Raychaudhuri equation, which reads
\begin{equation}\label{Raychaudhuri_Eq}
R_{ab}\ell^{a}\ell^{b}=\kappa \Theta-(\Theta_{ab}\Theta^{ab}-\Theta ^{2})-(1/\sqrt{q})(d/d\lambda)(\sqrt{q}\Theta)~.                                                                                                                     \end{equation} 
Substitution of the above expression for $R_{ab}\ell^{a}\ell^{b}$ along with the expression for $\partial _{a}(\sqrt{q}\Pi^{a}_{~b}\delta \ell^{b})$ and the result that $\ell_{a}P^{a}=-2\Theta$ leads to the following expression for the variation of the null boundary term associated with the Einstein-Hilbert action,
\begin{align}\label{Final_Eq}
-\delta \left\{\frac{1}{8\pi G}\int _{\partial \mathcal{V}}d^{2}xd\lambda\sqrt{-g}\left(\Theta+\kappa\right)\right\}
&=-\frac{1}{16\pi G}\int _{\partial \mathcal{V}}d^{2}xd\lambda \sqrt{q} \left\{\kappa \Theta-(\Theta_{ab}\Theta^{ab}-\Theta ^{2})-\frac{1}{\sqrt{q}}\frac{d \sqrt{q}\Theta}{d\lambda}\right\}
\nonumber
\\
&-\frac{1}{16\pi G}\int _{\partial \mathcal{V}}d^{2}xd\lambda \sqrt{q} 
\Big[\left\{\Theta_{ab}\Theta ^{ab}-(\Theta+\kappa)\Theta\right\}
+\kappa P_{a} \ell^{a}+\frac{1}{\sqrt{q}}\frac{d}{d\lambda}\left(\kappa \sqrt{q}\right)\Big]
\nonumber
\\
&=\frac{1}{8\pi G}\int _{\partial \mathcal{V}}d^{2}xd\lambda \sqrt{q} \kappa \Theta+\frac{1}{16\pi G}\int_{\partial \partial \mathcal{V}} d^{2}x \sqrt{q}\left(\Theta-\kappa\right) \Big |_{\lambda _{1}}^{\lambda_{2}}~.
\end{align}
Thus, neglecting the boundary term, we obtain,
\begin{align}
-\delta \left\{\frac{1}{8\pi G}\int _{\partial \mathcal{V}}d^{2}xd\lambda\sqrt{q}\left(\Theta+\kappa\right)\right\}
=\int _{\partial \mathcal{V}}d^{2}xd\lambda~ \frac{\kappa}{2\pi} \frac{d}{d\lambda}\left(\frac{\sqrt{q}}{4G}\right)
=\int _{\partial \mathcal{V}}d^{2}x~ T\, ds
\end{align}
where $ds$ corresponds to the rate of change of entropy along the null generator, i.e., $ds=(ds/d\lambda)d\lambda$. One can also arrive at another interesting thermodynamic result, even if one keeps the total derivative term in the above analysis, such that, 
\begin{align}
\delta \left\{\frac{1}{8\pi G}\int _{\partial \mathcal{V}}d^{2}xd\lambda\sqrt{q}\left(\Theta+\kappa\right)\right\}
&=-\int _{\partial \mathcal{V}}d^{2}xd\lambda~ \frac{\kappa}{2\pi} \frac{d}{d\lambda}\left(\frac{\sqrt{q}}{4G}\right)
+\frac{1}{16\pi G}\int_{\partial \partial \mathcal{V}} d^{2}x \sqrt{q}\left(\kappa-\Theta\right) \Big |_{\lambda _{1}}^{\lambda_{2}}
\nonumber
\\
&=\int _{\partial \mathcal{V}}d^{2}xd\lambda \frac{\sqrt{q}}{4G} \frac{d}{d\lambda}\left(\frac{\kappa}{2\pi}\right)
-\frac{1}{16\pi G}\int_{\partial \partial \mathcal{V}} d^{2}x \sqrt{q}\left(\kappa+\Theta\right) \Big |_{\lambda _{1}}^{\lambda_{2}}
=\int _{\partial \mathcal{V}}d^{2}x~ s\, dT
\end{align}
where, in the last line we have neglected the boundary contribution. We thus see that not only the boundary term but also its variation has a simple thermodynamic interpretation. Interestingly, the above result can also be arrived at from the first principle, which we have presented in \ref{AppC} for completeness.

\section{Conclusions}

Three peculiar features of gravitational theories which distinguishes them from other field theories are the following: 

(i) Gravity affects the propagation of light rays and hence the causal connection between events in spacetime; no other interaction is capable of doing this. A collective manifestation of this phenomena is exhibited by the null surfaces which can act as one way membranes for a particular class of observers. The observers who perceive the null surface as a horizon, limiting their vision, attributes to it a  heat density $Ts$. 

(ii) The most natural action principle for gravity contain second derivative of the dynamical variable. Integrating out these second derivatives lead to a surface term in the action and one needs to do something special to handle this surface term in order to obtain a sensible variational principle. The existence of such a surface term is yet another peculiar feature of gravity and is not prevalent in other field theories. 

(iii) Gravitational field equations can be interpreted entirely in a purely thermodynamic language. The heat density of null surfaces plays a crucial role in such a formulation. 

In this paper we have provided a simple synthesis of the three peculiar features of gravity listed above. We consider the action principle defined in a region with a boundary $\partial \mathcal{V}$ which could be either completely or partially a null surface. We then evaluate the boundary term on this null surface and show that it has a simple physical interpretation as the heat density of the null surface. We also show that the variations  induced by the flow along the null surface can be interpreted as thermodynamic variations, viz., $T\, \delta s$. These results reinforce the already well established idea that gravitational dynamics should be thought of as an emergent phenomena like fluid mechanics or elasticity.
\section*{Acknowledgements}

Research of S.C. is funded by the INSPIRE Faculty Fellowship (Reg. No. DST/INSPIRE/04/2018/000893) from Department of Science and Technology, Government of India. Research of TP is partially supported by the J.C.Bose Fellowship of the Department of Science and Technology, Government of India. Both the authors thank Bibhas Majhi and Krishnamohan Parattu for their comments on an earlier version of the manuscript. 
 
\appendix
\labelformat{section}{Appendix #1} 
\labelformat{subsection}{Appendix #1}
\section{Connection with spacelike/timelike surfaces}\label{AppA}

We have explicitly demonstrated that the term $s_{c}N^{c}_{ab}f^{ab}$ is related to $2\sqrt{q}\left(\Theta+\kappa\right)$ for null surfaces. Thus one may ask what happens for non-null surfaces, i.e. can we relate this to the boundary term of non-null surfaces, namely $2K\sqrt{h}$. We will demonstrate that we indeed can. To see this let us start with the boundary term in the gravitational action (no variation is involved), which reads,
\begin{align}
s_{c}N^{c}_{ab}f^{ab}=2\sqrt{-g}\Pi^{a}_{~b}\nabla_{a}s^{b}+\sqrt{-g}\left(-g^{ab}\partial_{a}s_{b}-\partial _{a}s^{a}\right)
-2\sqrt{-g}t^{a}s_{b}\nabla_{a}s^{b}
\end{align}
If we consider a spacelike surface (e.g., a $t=\textrm{constant}$ surface) then we have, $s_{c}=-n_{c}/N$, where $n_{c}$ is the normal and $N$ is the normalization factor. Thus we also have $t_{c}=-Nn_{c}$, such that $s_{c}t^{c}=-1$, since $n_{c}n^{c}=-1$. Thus the above expression can be expressed entirely in terms of the normal $n_{c}$, which yields,
\begin{align}
s_{c}N^{c}_{ab}f^{ab}&=2\sqrt{-g}\left(\delta^{a}_{b}+n^{a}n_{b}\right)\nabla_{a}\left(-\frac{1}{N}n^{b}\right)
+\sqrt{-g}\left[g^{ab}\partial_{a}\left(\frac{1}{N}n_{b}\right)+\partial_{b}\left(\frac{1}{N}n^{b}\right) \right]
\nonumber
\\
&+2\sqrt{-g}n^{a}n_{b}\nabla_{a}\left(\frac{1}{N}n^{b}\right)
\nonumber
\\
&=-\frac{2}{N}\sqrt{-g}h^{a}_{b}\nabla_{a}n^{b}+\sqrt{-g}\left[g^{ab}\partial_{a}\left(\frac{1}{N}n_{b}\right)+\partial_{b}\left(\frac{1}{N}n^{b}\right) \right]
\nonumber
\\
&+2\sqrt{-g}n^{a}n_{b}\nabla_{a}\left(\frac{1}{N}n^{b}\right)
\nonumber
\\
&=-2K\sqrt{h}+\sqrt{h}\left[g^{ab}\partial_{a}\left(n_{b}\right)+\partial_{b}\left(n^{b}\right) \right]
-2\sqrt{h}n^{a}\partial_{a}\ln N+2\sqrt{h}n^{a}\partial_{a}\ln N
\nonumber
\\
&=-2K\sqrt{h}+\sqrt{h}\left[g^{ab}\partial_{a}\left(n_{b}\right)+\partial_{b}\left(n^{b}\right) \right]
\end{align}
where we have used the fact that $n^{a}n^{b}\nabla_{a}n_{b}=0$, since the normal $n_{a}$ is unit normalized. Let us evaluate the second term in the above expression, which reads,
\begin{align}
g^{ab}\partial_{a}n_{b}+\partial_{b}n^{b}&=g^{tt}\partial_{t}n_{t}+g^{\alpha t}\partial_{\alpha}n_{t}
\nonumber
\\
&=-\frac{1}{N^{2}}\partial_{t}\left(-N\right)+\frac{N^{\alpha}}{N^{2}}\partial_{\alpha}\left(-N\right)
\end{align}
Thus if one moves to the synchronous frame such that $N=1$ and $N^{\alpha}=0$, then it immediately follows that the above term will vanish and hence the quantity $s_{c}N^{c}_{ab}f^{ab}$ becomes $-2K\sqrt{h}$, the boundary term associated with the spacelike surface. 
\section{Brief introduction to Gaussian Null Coordinates}\label{AppGNC}

In this section we will briefly describe  the Gaussian Null Coordinate system (often referred to as GNC), since many of our results fits in  naturally with this coordinate system. In flat spacetime, the null planes $X=\pm T$ are viewed as the horizon by comoving observers in Rindler coordinates. The GNC generalizes the notion of Rindler coordinate system to an \emph{arbitrary} null surface. The line element in this system of coordinates takes the following form:
\begin{align}
ds^{2}=-2r\alpha du^{2}+2dudr-2r\beta_{A}dudx^{A}+q_{AB}dx^{A}dx^{B}~.
\end{align}
The above metric is characterized by six unknown functions, $\alpha$, $\beta_{A}$ and $q_{AB}$ respectively, all of which are functions of $(u,r,x^{A})$. The surfaces $u=\textrm{constant}$ as well as the surface $r=0$ are null surfaces, while all the other $r=\textrm{constant}$ surfaces are spacelike. The Rindler metric, central to flat spacetime thermodynamics, is just a special case of the above metric, with $\alpha=\textrm{constant}$, $\beta_{A}=0$ and $q_{AB}=\delta_{AB}$. Thus any result presented using GNC will hold for \emph{any} class of null surfaces, which includes the Rindler and black hole horizons as special cases. 

The surface $r=0$ is the one of special interest, having the following normal vector $\ell_{a}=\nabla_{a}r$, such that $\ell_{a}\ell^{a}|_{r=0}=0$. For this null vector we have $\alpha(u,r,x^{A})$ as the non-affinity parameter, since $\ell^{a}\nabla_{a}\ell^{b}|_{r=0}=\alpha \ell^{b}$. In addition we need an auxiliary null vector to uniquely characterize the null surface, which we have denoted as $k^{a}$ and has the form $-(\partial/\partial r)^{a}$ in the GNC. In this system of coordinates we have $\ell^{a}=(1,2r\alpha+r^{2}\beta^{2},r\beta^{A})$ and hence $\partial_{a}\ell^{a}|_{r=0}=2\alpha$, proportional to the non-affinity parameter associated with the null surface. This is one of the results used in the main text. 

The intrinsic geometry associated with the $r=0$ null surface is characterized by the two-metric $q_{AB}$ and the null normal $\ell^{a}$; being null, it is also tangential to the null surface. On the null surface one can express, $\ell^{a}=(\partial/\partial u)^{a}$ and hence the null surface can be parameterized by the coordinates $(u,x^{A})$. The extrinsic curvature associated with the null surface is given by $\Theta_{AB}=(1/2)\partial _{u}q_{AB}$ and the expansion, i.e., the trace of the extrinsic curvature is give by $\Theta=\partial _{u}\ln \sqrt{q}$. Thus integral of $\Theta$ on the null surface will contribute only at the $u=\textrm{constant}$ boundaries of the null surface, which  can be ignored in bulk integrations by suitable boundary conditions. These are some of the results we have used in the main text.      
\section{Variation of the null boundary term}\label{AppB}

In this appendix we will present the structure of the variation of the boundary term. First of all, from the discussion around \ref{Therm_05} we have the following expression for the boundary term,
\begin{align}\label{Therm_05a}
\int _{\partial \mathcal{V}}d^{3}x~s_{c}f^{ab}\delta N^{c}_{ab}=-\int _{\mathcal{V}}d^{4}x f^{ab}\delta R_{ab}~.
\end{align}
Concentrating on the integrand of the boundary term appearing on the left hand side of the above expression we obtain,
\begin{align}\label{Therm_06}
s_{c}f^{ab}\delta N^{c}_{ab}&=\sqrt{-g}s_{c}\nabla _{d}\left(\delta g^{cd}-g^{cd}g_{ik}\delta g^{ik}\right)
\nonumber
\\
&=\sqrt{-g}\nabla _{d}\left(s_{c}\delta g^{cd}\right)-\sqrt{-g}\left(\nabla_{d}s_{c}\right)\delta g^{cd}-\sqrt{-g}s^{c}\nabla _{c}\left(g_{ik}\delta g^{ik}\right)
\nonumber
\\
&=\sqrt{-g}\nabla _{d}\left(\delta s^{d}\right)-\sqrt{-g}\left(\nabla_{c}s_{d}\right)\delta g^{cd}+2\sqrt{-g}s^{c}\nabla _{c}\delta \ln \sqrt{-g}~,
\end{align}
where, we have assumed that under variation the boundary surface $\phi=\textrm{constant}$ does not change, such that $\delta s_{c}=0$. We can now use the following relation, $\delta (\nabla_{c}s^{c})=\nabla_{c}\delta s^{c}+s^{c}\nabla_{c}\delta \ln \sqrt{-g}$ to rewrite \ref{Therm_06} as,
\begin{align}\label{Therm_07}
s_{c}f^{ab}\delta N^{c}_{ab}&=-\sqrt{-g}\nabla _{d}\left(\delta s^{d}\right)-\sqrt{-g}\nabla_{c}s_{d}\delta g^{cd}
+2\sqrt{-g}\delta \left(\nabla_{c}s^{c}\right)~.
\end{align}
Hence the boundary term in the variation of the Einstein-Hilbert action becomes,
\begin{align}\label{Therm_08}
\mathcal{B}&\equiv -\int _{\partial \mathcal{V}}d^{3}x~s_{c}f^{ab}\delta N^{c}_{ab}
\nonumber
\\
&=\int _{\partial \mathcal{V}}d^{3}x~\left[\sqrt{-g}\nabla _{d}\left(\delta s^{d}\right)+\sqrt{-g}\nabla_{c}s_{d}\delta g^{cd}
-2\sqrt{-g}\delta \left(\nabla_{c}s^{c}\right)\right]~.
\end{align}
This expression holds true for spacelike/timelike as well as for null surfaces. For spacelike/timelike surfaces, we can choose, the normal to be $n_{a}=-Ns_{a}$, where $n_{a}n^{a}=\epsilon$ and $\epsilon=\mp 1$ denotes spacelike/timelike surfaces. Then the above expression for the boundary term in the variation of the gravitational action yields \cite{Parattu:2016trq},
\begin{align}\label{Therm_09}
\mathcal{B}=\int _{\partial \mathcal{V}}d^{3}x\left[\sqrt{h}\left(D_{a}\delta n^{a}\right)-\delta \left(2K\sqrt{h}\right)
+\sqrt{h}\left(K_{ab}-Kh_{ab}\right)\delta h^{ab}\right]~.
\end{align}
However, our main interest is in the context of null hypersurfaces, where the null normal is denoted by $\ell_{a}=\nabla_{a}\phi$, with the variation satisfying the following conditions, $\delta \ell_{a}=0$, $\delta (\ell_{a}\ell^{a})=0=\delta(\ell_{a}k^{a})$. Thus from the general result, presented in \ref{Therm_08}, it immediately follows that the boundary term in the variation of the Einstein-Hilbert action for null surfaces read, 
\begin{align}\label{Therm_10}
\mathcal{B}&=\int _{\partial \mathcal{V}}d^{3}x\Bigg[\partial _{a}\left(\sqrt{q}\Pi^{a}_{b}\delta \ell^{b}\right)
-2\delta\left\{\sqrt{-g}\left(\Theta+\kappa\right)\right\}+\sqrt{q}\left\{\Theta_{ab}-(\Theta+\kappa)q_{ab}\right\}\delta q^{ab}
\nonumber
\\
&+\sqrt{q}\left\{2k_{a}\left(\Theta+\kappa\right)-k^{b}\left(\nabla_{a}\ell_{b}+\nabla_{b}\ell_{a}\right) \right\}\delta \ell^{a}\Big]~.
\end{align}
Here the symbols have their usual meaning. Finally using \ref{Therm_10} and \ref{Therm_05a} one can obtain an expression for the variation of the null boundary term, $2\sqrt{-g}\left(\Theta+\kappa\right)$ in terms of $\delta R_{ab}$, $\delta q^{ab}$ and $\delta \ell^{a}$, which is used in the main text. 

As an aside,  we make the following clarification about the variations of the kind $\delta g^{ab}=(1/2)(\nabla^{a}\ell^{b}+\nabla^{b}\ell^{a})$ arising from the flow $x^{a}\to x^{a}+(1/2)\ell^{a}$. In such variations one should think of $\ell^a$ as $\epsilon\ell^a$ and one takes the $\epsilon\to0$ limit at an appropriate juncture. In addition to making the infinitesimal nature of variations well-defined, the change $\ell^a\to\epsilon\ell^a$ is also needed for dimensional reasons --- a fact which is sometimes not appreciated. In general, metric coefficients can have any dimension depending on the coordinate choice. If we choose $\ell_{a}=\nabla_{a}\phi$, it is dimensionless since $\phi$ can be treated as a coordinate with dimension of length. So the $\epsilon$ is required to match dimensions in both sides of the equation $\delta g^{ab}=(\epsilon/2)(\nabla^{a}\ell^{b}+\nabla^{b}\ell^{a})$. We do not exhibit $\epsilon$ explicitly since we get the same result by that route as well.
\section{Another perspective in the variation of null boundary term}\label{AppC}

It is also possible to provide yet another perspective to the variation of the null boundary term. In this we will assume that $\delta \ell_{a}=0$, leading to $\delta \ell^{a}=\kappa \ell^{a}$ as $\delta g^{ab}=(1/2)(\nabla^{a}\ell^{b}+\nabla^{b}\ell^{a})$. This ensures that $\delta \ell^{2}=0$, as desired. Similarly, we need to impose the condition $\delta \left(\ell^{a}k_{a}\right)=0$, which yields, $\ell^{a}\delta k_{a}=\kappa$. Thus we should not assume $\delta k_{a}=0$. 

We have derived the result that $\delta \left\{2\sqrt{-g}\left(\Theta+\kappa\right) \right\}$ for the above variation of the metric equals to $Tds$, using the variation of the gravitational action. However, the above result is also derivable from direct variation of the null boundary term as well. For this purpose we need to compute the variation of the surface gravity, which can be achieved by considering variation of the equation defining surface gravity. This yields,
\begin{align}
\ell_{b}\delta \kappa&=\delta \left(\ell^{a}\nabla_{a}\ell_{b}\right)
=\nabla_{a}\ell_{b}\delta \ell^{a}+\ell^{a}\delta \left(\nabla_{a}\ell_{b}\right)
\nonumber
\\
&=\kappa \ell^{a}\nabla_{a}\ell_{b}+\ell^{a}\ell_{c}\left(-\delta \Gamma^{c}_{ab}\right)
\nonumber
\\
&=\kappa^{2}\ell_{b}-\frac{1}{2}\ell^{a}\ell^{c}\left(-\nabla_{c}\delta g_{ab}+\nabla_{a}\delta g_{bc}+\nabla_{b}\delta g_{ac}\right)~.
\end{align}
Thus taking inner product of this equation with $k^{b}$ we finally arrived at,
\begin{align}
\delta \kappa&=\kappa^{2}+\frac{1}{2}k^{b}\ell^{a}\ell^{c}\left(-\nabla_{c}\delta g_{ab}+\nabla_{a}\delta g_{bc}+\nabla_{b}\delta g_{ac}\right)
=\kappa^{2}+\frac{1}{2}k^{b}\ell^{a}\ell^{c}\nabla_{b}\delta g_{ac}
\nonumber
\\
&=\kappa^{2}-\frac{1}{2}k^{b}\ell^{a}\ell^{c}\nabla_{b}\nabla_{a}\ell_{c}
=\kappa^{2}-\frac{1}{2}k^{b}\nabla_{b}\left(\ell^{a}\ell^{c}\nabla_{a}\ell_{c}\right)+\frac{1}{2}k^{b}\nabla_{a}\ell_{c}\nabla_{b}\left(\ell^{a}\ell^{c}\right)
\nonumber
\\
&=\kappa^{2}-\frac{1}{2}k^{b}\nabla_{b}\left(\frac{1}{2}\ell^{a}\nabla_{a}\ell^{2}\right)-\kappa^{2}~.
\end{align}
Even though it is tempting to set $(1/2)\ell^{a}\nabla_{a}\ell^{2}=0$, one should note that this holds only on the null surface, while the above expression involves off the null surface derivative as well. Specializing to the GNC construction \cite{Parattu:2015gga} we see,
\begin{align}
\delta \kappa&=-\frac{1}{4}k^{r}\partial_{r}\left(\ell^{u}\partial_{u}\ell^{2}+\ell^{r}\partial_{r}\ell^{2}\right)
=\frac{1}{4}\partial_{r}\left[\partial_{u}\left(2r\kappa\right)+2r\kappa \partial_{r}\left(2r\kappa\right)\right]
\nonumber
\\
&=\frac{1}{4}\partial_{r}\left[2r\partial_{u}\kappa+4r\kappa^{2}\right]
=\kappa^{2}+\frac{1}{2}\ell^{a}\nabla_{a}\kappa~.
\end{align}
One can also start from the alternative expression for $\kappa$, which is $-k^{a}\ell^{b}\nabla_{b}\ell_{a}$ and vary it. Since $\delta \left(\ell^{a}k_{a}\right)=0$, this will lead to an identical expression as above. At this stage, one point should be emphasized, namely the variation itself carries a dimension. This is because, the metric is assumed to be dimensionless, so is $\ell_{a}$. Thus the covariant derivative of $\ell^{a}$ and hence $\delta g^{ab}$ must have dimension of inverse length. This is why variation of $\kappa$ involves a $\kappa^{2}$ term. 

As the second part of the variation of the null boundary term for Einstein-Hilbert action let us compute the variation of the expansion scalar associated with the null generator, yielding, 
\begin{align}
\delta \Theta&=\delta \left(q^{ab}\nabla_{a}\ell_{b}\right)
=\delta q^{ab} \left(\nabla_{a}\ell_{b}\right)+q^{ab} \delta \left(\nabla_{a}\ell_{b}\right)
\nonumber
\\
&=\left(\nabla_{a}\ell_{b}\right) \left(\delta g^{ab}+k^{b}\delta \ell^{a}+\ell^{a}\delta k^{b}+\ell^{b}\delta k^{a}+k^{a}\delta \ell^{b}\right)
-\frac{1}{2}q^{ab}\ell^{c}\left(-\nabla_{c}\delta g_{ab}+\nabla_{a}\delta g_{bc}+\nabla_{b}\delta g_{ac}\right)
\nonumber
\\
&=\nabla_{a}\ell_{b}\nabla^{a}\ell^{b}+2\kappa \ell_{a}\delta k^{a}-2\kappa^{2}
-\frac{1}{2}q^{ab}\ell^{c}\nabla_{c}\nabla_{a}\ell_{b}+q^{ab}\ell^{c}\nabla_{a}\nabla_{b}\ell_{c}
\nonumber
\\
&=\Theta_{ab}\Theta^{ab}-\frac{1}{2}q^{ab}\ell^{c}\left[\nabla_{c},\nabla_{a}\right]\ell_{b}-\frac{1}{2}q^{ab}\ell^{c}\nabla_{a}\nabla_{c}\ell_{b}+q^{ab}\ell^{c}\nabla_{a}\nabla_{b}\ell_{c}
\nonumber
\\
&=\Theta_{ab}\Theta^{ab}+\frac{1}{2}q^{ab}\ell^{c}R^{p}_{~bca}\ell_{p}-\frac{1}{2}q^{ab}\nabla_{a}\left(\ell^{c}\nabla_{c}\ell_{b}\right)+q^{ab}\nabla_{a}\left(\ell^{c}\nabla_{b}\ell_{c}\right)
-\frac{1}{2}q^{ab}\left(\nabla_{a}\ell^{c}\right)\left(\nabla_{b}\ell_{c}\right)
\nonumber
\\
&=\Theta_{ab}\Theta^{ab}+\frac{1}{2}R_{ab}\ell^{a}\ell_{b}+\frac{1}{2}\kappa \Theta-\frac{1}{2}\Theta_{ab}\Theta^{ab} 
\nonumber
\\
&=\frac{1}{2}R_{ab}\ell^{a}\ell_{b}+\frac{1}{2}\kappa \Theta+\frac{1}{2}\Theta_{ab}\Theta^{ab}~.
\end{align}
Thus combining the variation of the non-affinity parameter and the expansion scalar we finally obtain the following expression for the variation of the null boundary term,
\begin{align}
\delta \left[2\sqrt{-g}\left(\Theta+\kappa\right)\right]&=2\sqrt{-g}\left(\delta\Theta+\delta\kappa\right)+2\left(\Theta+\kappa\right)\delta \sqrt{-g}
\nonumber
\\
&=2\sqrt{-g}\left(\frac{1}{2}R_{ab}\ell^{a}\ell_{b}+\frac{1}{2}\kappa \Theta+\frac{1}{2}\Theta_{ab}\Theta^{ab}+\kappa^{2}+\frac{1}{2}\ell^{a}\nabla_{a}\kappa\right)+2\left(\Theta+\kappa\right)\left(-\frac{1}{2}\sqrt{-g}g_{ab}\delta g^{ab} \right)
\nonumber
\\
&=2\sqrt{-g}\left(\frac{1}{2}R_{ab}\ell^{a}\ell_{b}+\frac{1}{2}\kappa \Theta+\frac{1}{2}\Theta_{ab}\Theta^{ab}+\kappa^{2}+\frac{1}{2}\ell^{a}\nabla_{a}\kappa\right)-\sqrt{-g}\left(\Theta+\kappa\right)\nabla_{a}\ell^{a}
\nonumber
\\
&=\sqrt{-g}\left(R_{ab}\ell^{a}\ell_{b}+\kappa \Theta+\Theta_{ab}\Theta^{ab}+2\kappa^{2}+\ell^{a}\nabla_{a}\kappa\right)-\sqrt{-g}\left(\Theta+\kappa\right)\left(\Theta+2\kappa\right)
\nonumber
\\
&=\sqrt{-g}\left(R_{ab}\ell^{a}\ell_{b}-3\kappa \Theta+\Theta_{ab}\Theta^{ab}-\Theta^{2}+\ell^{a}\nabla_{a}\kappa\right)
\end{align}
Using the Raychaudhuri equation presented in \ref{Raychaudhuri_Eq}, we obtain,
\begin{align}
-\delta \left[2\sqrt{-g}\left(\Theta+\kappa\right)\right]&=\sqrt{-g}\left(2\kappa \Theta\right)-\left(\kappa \Theta+\ell^{a}\nabla_{a}\kappa\right)+\frac{1}{\sqrt{q}}\frac{d\sqrt{q}\Theta}{d\lambda}
\nonumber
\\
&=\sqrt{q}\left(2\kappa \Theta\right)+\frac{1}{\sqrt{q}}\frac{d}{d\lambda}\left\{\sqrt{q}\left(\Theta-\kappa \right)\right\}
\end{align}
Thus dividing the above expression by $16\pi G$ we again get back the desired expression presented in \ref{Final_Eq}. 

\bibliography{bibliography}

\bibliographystyle{./utphys1}
\end{document}